\documentclass[preprints,article,accept,pdftex,moreauthors]{Definitions/mdpi}

\firstpage{1} 
\pubvolume{12}
\issuenum{7}
\articlenumber{3594}
\pubyear{2022}
\copyrightyear{2022}
\externaleditor{Academic Editor: {Luis Javier García Villalba} %
} %
\datereceived{09 March 2022} 
\dateaccepted{29 March 2022} 
\datepublished{1 April 2022} 
\hreflink{https://doi.org/10.3390/app12073594} %
\doinum{10.3390/app12073594}
\pdfoutput=1

\newcolumntype{C}[1]{>{\PreserveBackslash\centering}m{#1}}
\newcolumntype{R}[1]{>{\PreserveBackslash\raggedleft}m{#1}}
\newcolumntype{L}[1]{>{\PreserveBackslash\raggedright}m{#1}}

\usepackage[labelformat=simple]{subcaption}

\DeclareCaptionLabelFormat{subcaptionlabel}{\normalfont(\textbf{#2}\normalfont)}
\usepackage{multirow}
\usepackage{graphicx}

\usepackage{bm}
\usepackage{amsfonts}

\usepackage{bbding}
\usepackage{pifont}
\usepackage{wasysym}

\usepackage[whole]{bxcjkjatype} %

\Title{Ad Creative Discontinuation Prediction with Multi-Modal Multi-Task Neural Survival Networks}

\TitleCitation{Ad Creative Discontinuation Prediction with Multi-Modal Multi-Task Neural Survival Networks}

\Author{Shunsuke Kitada %
 $^{1,}$*$^{,\dagger}$\orcidA{}, Hitoshi Iyatomi $^{1}$\orcidB{} and Yoshifumi Seki $^{2}$}

\AuthorNames{Shunsuke Kitada, Hitoshi Iyatomi and Yoshifumi Seki}

\AuthorCitation{Kitada, S.; Iyatomi, H.; Seki, Y.}

\address{%
$^{1}$ \quad Department of Applied Informatics, Graduate School of Science and Engineering, Hosei University, Tokyo, 184-8584, Japan; shunsuke.kitada.8y@stu.hosei.ac.jp (S.K.), iyatomi@hosei.ac.jp (H.I.) \\ %
$^{2}$ \quad Gunosy Inc., Tokyo, 150-6139, Japan; yoshifumi.seki@gunosy.com}

\corres{Correspondence: shunsuke.kitada.8y@stu.hosei.ac.jp}

\firstnote{This work was conducted during the first author's internship at Gunosy Inc.}

\abstract{Discontinuing ad creatives at an appropriate time is one of the most important ad operations that can have a significant impact on sales. Such operational support for ineffective ads has been less explored than that for effective ads. After pre-analyzing 1,000,000 real-world ad creatives, we found that there are two types of discontinuation: short-term (i.e., cut-out) and long-term (i.e., wear-out). In this paper, we propose a practical prediction framework for the discontinuation of ad creatives with a hazard function-based loss function inspired by survival analysis. Our framework predicts the discontinuations with a multi-modal deep neural network that takes as input the ad creative (e.g., text, categorical, image, numerical features). To improve the prediction performance for the two different types of discontinuations and for the ad creatives that contribute to sales, we introduce two new techniques: (1) a two-term estimation technique with multi-task learning and (2) a click-through rate-weighting technique for the loss function. We evaluated our framework using the large-scale ad creative dataset, including 10 billion scale impressions. In terms of the concordance index (short: 0.896, long: 0.939, and overall: 0.792), our framework achieved significantly better performance than the conventional method (0.531). Additionally, we confirmed that our framework (i) demonstrated the same degree of discontinuation effect as manual operations for short-term cases, and (ii) accurately predicted the ad discontinuation order, which is important for long-running ad creatives for long-term cases.
}


\featuredapplication{This work can be applied to support decisions to discontinue ad creatives, which is expected to reduce the load of ad operators. Currently, our framework is deployed in a production environment and provides predictions for our internal operators. Based on the predictions, we have discussions with the operations team to improve the operation efficiency.}

\begin{document}

\section{Introduction}\label{sec:intro}
With the increasing importance of digital advertising, the significance of ad operations has also increased.
Ad operations consist of various tasks, such as preparing ad creatives, determining the bid price and the target audience, and discontinuing inefficient ad creatives.
These operations are essential for business revenue; however, operators' abilities are dependent on their experience.
Supporting ad operations is an important topic of study in the field of machine learning (ML)  \cite{kitada19, mishra2019guiding, maehara2018optimal, yang2019bid, Doan2019adversarial}.
Discontinuing ad creatives at the appropriate time is a crucial operation and can have a significant impact on sales. However, to the best of our knowledge, no studies have been conducted so far that support these operations.

To support ad operations, we address the discontinuation of ineffective ad creatives based on ML methods.
Ad creatives are generally discontinued manually by ad operators when they become less effective.
Considering that ad-serving algorithms (e.g., bid price optimization \cite{maehara2018optimal, yang2019bid}) are required to serve a large number of effective ad creatives, multiple studies have been conducted to develop algorithms that can provide accurate \mbox{predictions \cite{shatnawi2012statistical, chen2016deep}.}
However, we posit that predicting ad performance for operational efficiency requires other mechanisms because the utilization of the results is different from efficient ad serving.
In other words, in this study, we focus on predicting the effectiveness of the ad creatives so as to properly discontinue ineffective ones.

Pre-analyzing 1,000,000 real-world ad creatives, we found two types/patterns of discontinuation: short-term and long-term. 
These discontinuations were empirically identified as falling into two categories: (a) \textit{cut-out} (i.e., for short-term discontinuation) and (b) \textit{wear-out} (i.e., for long-term discontinuation).
In cut-out discontinuation, ad creatives are discontinued after a short period of time due to incompatible serving effects, whereas in wear-out discontinuation, ad creatives are discontinued when ad creatives' performance deteriorates after a long time.
Ad operators need to take different actions for each type of discontinuation.
We clarified these empirical differences by analyzing real-world data (described in detail in Section {\ref{sec:dataset_overview}}).

To address the discontinuation of the ad creatives that have an impact on business revenue, we propose a discontinuation prediction framework based on the real-world data; this is the first piece of work focusing on ineffective ad creatives, rather than effective ones. Our framework is a multi-modal deep neural network (DNN) with multiple ad creative features (e.g., text, categorical, image, numerical features) as inputs and incorporates a hazard function, which is frequently used in survival prediction, into a loss function. (It is controversial to solve ad discontinuation prediction as a survival prediction. In this paper, we do not elaborate on this topic but only discuss the idea of the hazard function. We intend to tackle this issue in our future studies.)
To improve the prediction performance for the two different types of discontinuations and for the ad creatives that contribute to sales, we introduce two new techniques: (1) a two-term estimation technique with multi-task learning (MTL) \cite{caruana1997multitask} and (2) a click-through rate (CTR)-weighting technique for the loss function (described in detail in Section {\ref{sec:method}}).
Using the large-scale ad creative dataset, we practically evaluated the proposed framework and techniques with various metrics such as concordance index (CI) \cite{harrell1982evaluating}, the F1 score, and operational performance measures (described in detail in Sections~\ref{sec:offline} and \ref{sec:online}).

The contributions of our study are summarized as follows:
\begin{itemize}
 \item \textbf{Analyzing large-scale ad creative dataset}: We analyzed 1,000,000 real-world ad creatives, including 10 billion scale impressions, and identified two important aspects. (a) There are two reasons for the discontinuation of ad creatives: cut-out (for the short term) and wear-out (for the long term). (b) The business impact of discontinuation varies significantly for different ad creatives.
 \item \textbf{Proposing a new framework for predicting ad creative discontinuation}: We propose a multi-modal DNN-based framework for predicting ad discontinuation with a hazard function-based loss function. Our framework includes two effective techniques: (i) a two-term estimation technique with MTL and (ii) a CTR-weighting technique for the loss function. Our framework demonstrated significantly better performance (short: 0.896, long: 0.939, and overall: 0.792 in the CI) than the conventional method (0.531 in the CI) did. Additionally, our framework outperformed classification- and regression-based frameworks.
 \item \textbf{Evaluating the framework through practical case studies}: We evaluated our framework from two important perspectives of case studies and confirmed the following: For short-term cases, our framework demonstrated an equivalent or better discontinuation effect compared with manual operations. For long-term cases, our framework provided a high degree of similarity of order to manual discontinuation orders compared with other methods, based on other practical indicators; this is important for long-running ad creatives.
\end{itemize}

The remainder of this paper is organized as follows.
Section~\ref{sec:related_work} reviews related work about supporting ad operations and survival prediction/hazard functions.
Section~\ref{sec:dataset_overview} describes the insight from the pre-data analysis and clarifies the characteristics of each type of discontinuation.
Section~\ref{sec:method} presents the multi-modal DNN-based framework for predicting the timing of the discontinuation of ad creatives.
Section~\ref{sec:offline} describes offline experiments with standard metrics using already-observed (i.e., offline) data.
Section~\ref{sec:online} describes case studies that simulate online experiments using as-yet-unobserved (i.e., online) data.
Section~\ref{sec:limitation} describes the limitations and future works of this study.
Finally, Section~\ref{sec:conclusion} concludes this study.

\section{Related Work}\label{sec:related_work}

In this section, we describe related work about supporting ad creative operation process and survival prediction/hazard functions.

\subsection{Supporting the Ad Creative Operation Processes}

Table \ref{tab:related_work} shows a classification of ML methods that support the process for ad creatives.
The operation process for ad creatives generally includes the following: (1) created and submitted to the ad platform, (2) served to users on the platform, and (3) discontinued when they become less effective.
There are several studies to support (1) and (2) based on ML methods, such as supporting the creation of high-performing ad \mbox{creatives \cite{Thomaidou2013automated, kitada19, mishra2019guiding, hughes2019generating},} determining the best bid price \cite{zhang2012joint, maehara2018optimal, Ren2018Bidding}, and deciding to whom to serve the ad creatives \cite{shen2016effective, Liu2016audience, Doan2019adversarial}.
However, to the best of our knowledge, no studies have been conducted on supporting the discontinuation process.
Thus, we focus on process (3) and develop a framework for predicting the appropriate timing for the discontinuation of a ad creative.

The serving performance of ads can be inefficient for a variety of reasons, one of which is wear-out.
Wear-out is an event caused by decreasing ad performance due to repeated serving to users.
To control for wear-out, frequency capping, (\url{https://support.google.com/google-ads/answer/117579}; accessed on 31st March 2022) which restricts the number of times an ad creative is served to a user, is a traditional function, and the number of servings of an ad is useful in predicting the CTR and/or conversion rate (CVR) \cite{shatnawi2012statistical}.
The event is commonly discussed in marketing science literature \cite{cornelia1988advertising}, and an ad-serving algorithm that incorporates wear-out as a feature has been previously proposed \cite{morisaki2020fatigue}.
However, no studies have directly predicted the wear-out of ad creatives.

In related studies about ad creative discontinuation, a bandit algorithm has been commonly used for selecting effective ad creatives \cite{Tang2013automatic}.
Using the algorithm, the efficiency of ad creatives can be estimated correctly with fewer impressions.
However, ad operators must decide which ads to discontinue based on their performance; determining the appropriate timing of discontinuation demands much experience.

\begin{table}[H]
\caption{Classification of machine learning methods to support the process for ad creatives.}
\label{tab:related_work}
\begin{adjustwidth}{-\extralength}{0cm}
\begin{tabularx}{\fulllength}{ll}
\toprule
\textbf{The Process for Ad Creative Operations} & \textbf{Machine Learning Methods to Support the Operation Process} \\ 
\cmidrule(r){1-1} \cmidrule(l){2-2}
(1) Creating and submitting to the ad platform & Thomaidou et al. \cite{Thomaidou2013automated}, Kitada et al. \cite{kitada19}, Hughes et al. \cite{hughes2019generating}, Mishra et al. \cite{mishra2019guiding} \\
(2) Serving to users on the platform & Zhang et al. \cite{zhang2012joint}, Shen et al. \cite{shen2016effective}, Maehara et al. \cite{maehara2018optimal}, Doan et al. \cite{Doan2019adversarial} \\
(3) Discontinuing when the ads become less effective & \textbf{Our study} \\ \bottomrule
\end{tabularx}
\end{adjustwidth}
\end{table}

\subsection{Survival Prediction and Hazard Function}

Survival prediction is a branch of statistics that deals with time-to-event data, that is, data where the objective variable is the time until the event occurs.
The effectiveness of the strategy has been confirmed across various domains, such as credit score analysis \cite{dirick2017time} and customer analysis \cite{van2004customer}.
In digital advertising, survival prediction helps to predict the conversion rate \cite{chapelle2014modeling}, user experiences after clicking on an ad \cite{barbieri2016improving}, the winning bid price \cite{wu2015predicting}, and the failure rate of the reserve price \cite{kalra2019reserve}.

The main feature of the survival prediction task is to correctly model censored data such that no event occurs during the observation period.
However, in the ad discontinuation task, such data are scarce because the operator manually determines the time of discontinuation. (Although there are some cases where the serving volume is low, they are very few.)
When such censoring data are scant or nonexistent, Zhong et al. \cite{zhong2019survival} reported that the survival prediction task can be regarded as equivalent in terms of formulation to the classification task.
Although it is debatable whether survival prediction is suitable for ad discontinuation tasks, we assume that the properties of ad discontinuation have some similarities with those of survival prediction tasks.
Thus, we attempt to utilize the hazard function, which is an important idea in survival prediction.

The hazard function can model the main variable of interest as the time until an event.
This function has been used to represent the decaying survival probability, and deep learning-based survival prediction models have been developed by training a loss function based on the function \cite{katzman2018deepsurv, lee2018deephit, gensheimer2019scalable}.
We considered that the function is suitable for modeling the discontinuation probability of exponentially decaying ad creatives (described in detail in Section \ref{sec:dataset_overview}).
Through training based on the hazard function with a deep learning model, we expect to develop a better prediction model that will help ad operators with discontinuation.

\section{Pre-Data Analysis}\label{sec:dataset_overview}
In this section, we clarify the characteristics of the two types of discontinuation (cut-out and wear-out) by analyzing a real-world dataset.
First, we provide an overview of the dataset.
Second, we analyze the dataset in terms of serving days and sales aspects.
Then, we discuss the difference between the two types of discontinuation based on the analysis.

\subsection{Dataset Overview}

We used real-world data from the Japanese digital advertising program Gunosy Ads, (Gunosy Ads \url{https://gunosy.co.jp/ad/en/}; accessed on 31st March 2022) provided by Gunosy Inc. (Gunosy Inc. \url{https://gunosy.co.jp/en/}; accessed on 31st March 2022)
Gunosy Inc. is a provider of several news delivery applications, and Gunosy Ads delivers digital advertisements for these applications. 
The applications have been downloaded more than 53 million times as of November 2019.
We used approximately 1,000,000 ad creatives, including 10 billion scale impressions, served from 2018 to 2019 by Gunosy Ads.
The ads were managed in units called \textit{campaigns}, and each campaign was configured with a target gender, ad genre, and cost per acquisition (CPA).
Table {\ref{tab:dataset_samples}} shows the examples included in the ad creative dataset.
Each ad creative was associated with a campaign and consisted of texts (e.g., title, description), categorical information (e.g., gender of the serving target, genre of the ad creative), and an image attached to the ad creative, and numerical information (e.g., the number of impressions, clicks, conversions, and the CPA for the ad creative). 
Ad performance data, including the number of impressions, clicks, conversions, and sales, were tabulated for each ad creative daily. 
Because the data are confidential, the actual values cannot be disclosed. 
Thus, relative values are used when discussing the data.

\begin{table}[H]
\centering
\caption{Examples included in the ad creative dataset. In addition to the Japanese title text and description text shown here, the dataset contains categorical information (e.g., gender of the serving target, genre of the ad creative), images (e.g., the image attached to the ad creative), and numerical information (e.g., the number of impressions, clicks, conversions, and the CPA for the ad creative).}
\label{tab:dataset_samples}
\begin{adjustwidth}{-\extralength}{0cm}
\begin{tabularx}{\fulllength}{llcc}
\toprule
\multicolumn{1}{c}{\textbf{Title Text}} & \multicolumn{1}{c}{\textbf{Description Text}} & \textbf{Gender} & \textbf{Genre} \\ \cmidrule(r){1-1} \cmidrule(lr){2-2} \cmidrule(lr){3-3} \cmidrule(l){4-4}
\begin{tabular}[c]{@{}l@{}}{\footnotesize 1000万人が選ぶ！みんなが選んでいるゲーム10選}\\ \textit{\footnotesize Chosen by 10 million people! The 10 games played by everyone.}\end{tabular} & \begin{tabular}[c]{@{}l@{}}{\footnotesize スマホに入れておきたい無料ゲームを限定でご紹介}\\ \textit{\footnotesize Exclusively introducing free games} \\ \textit{\footnotesize that you will want to install on mobile phone.}\end{tabular} & {\footnotesize All} & {\footnotesize Game App.} \\ \cmidrule(r){1-1} \cmidrule(lr){2-2} \cmidrule(lr){3-3} \cmidrule(l){4-4}
\begin{tabular}[c]{@{}l@{}}{\footnotesize -10kgのダイエットに成功！痩せる理由はこれ}\\ \textit{\footnotesize Success in -10 kg weight loss! This is the reason for getting slim.}\end{tabular} & \begin{tabular}[c]{@{}l@{}}{\footnotesize 女子に人気の方法で効果を実感}\\ \textit{\footnotesize Realizing the effects popular among girls.}\end{tabular} & {\footnotesize Female} & {\footnotesize Healthy Food} \\ \cmidrule(r){1-1} \cmidrule(lr){2-2} \cmidrule(lr){3-3} \cmidrule(l){4-4}
\begin{tabular}[c]{@{}l@{}}{\footnotesize 有名芸能人監修。簡単にできる料理レシピ本}\\ \textit{\footnotesize Supervised by a famous celebrity; easy cookbook.}\end{tabular} & \begin{tabular}[c]{@{}l@{}}{\footnotesize 一人暮らしの男性にもおすすめ！}\\ \textit{\footnotesize Recommended for men living alone!}\end{tabular} & {\footnotesize Male} & {\footnotesize Books} \\ \bottomrule
\end{tabularx}
\end{adjustwidth}
\end{table}

\subsection{Analysis of the Dataset}\label{sec:analysis_of_the_dataset}

In this section, we report the results of pre-data analysis using the real-world ad creative dataset in terms of serving days and sales. Based on these results, we found that there are two types of ad discontinuation.

\subsubsection{Variation of Serving Days}
\label{sec:analysis_variation_of_continuous_days}

\begin{figure}[H]
 \centering
 \includegraphics[width=\linewidth]{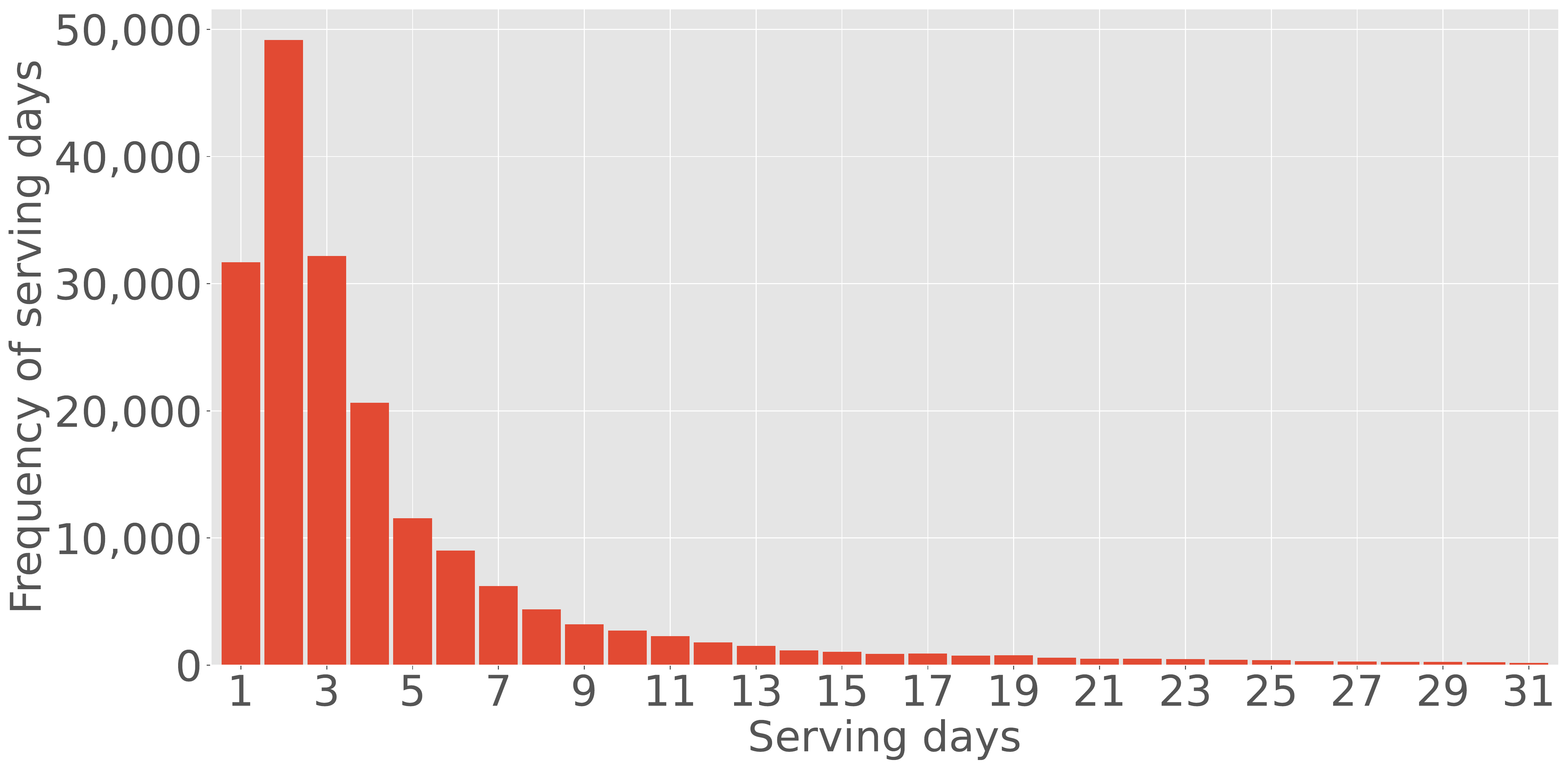}
 \caption{
 Frequency distribution of the number of serving days of the overall ad creatives in one month.
 Many creatives were discontinued within three days of serving.
 }
 \label{fig:freq_survival_day}
\end{figure}

First, we analyzed the distribution of the serving days, which is the period of time since the ad creatives were served, as in Figure \ref{fig:freq_survival_day}.
Many ad creatives were discontinued within three days of serving, and the continuous serving days of the ad creatives resemble an exponential distribution.

Table \ref{tab:serving_days_per_num_ads_and_sales} shows the percentage of the number of ads served and their sales on the serving days.
Although the number of ad creatives that were served for a long term is small, these creatives account for a large percentage of advertising sales.
Specifically, comparing the serving days of $[0, 3)$ and $[7, +\infty)$, the former has the majority of the ad creatives, but the latter accounts for the highest percentage of sales.
From this difference, we assume that each discontinuation is caused by different operations.
We discuss the differences in detail in Section \ref{sec:analysis_two_types_of_the_discontinuation}.

\begin{table}[H]
\caption{
Percentage of the number of ads served and their sales on the serving date.
While 80\% of the ads were discontinued within a week of serving, the 20\% of the ads that were served for more than a week accounted for 80\% of the sales.
}
\label{tab:serving_days_per_num_ads_and_sales}
\newcolumntype{C}{>{\centering\arraybackslash}X}
\begin{tabularx}{\textwidth}{Crr}
\toprule
\multicolumn{1}{c}{\textbf{Serving Days}} & \multicolumn{1}{l}{\textbf{Number of Ads} {[}\%{]}} & \multicolumn{1}{l}{\textbf{Sales} {[}\%{]}} \\ \cmidrule(r){1-1} \cmidrule(lr){2-2} \cmidrule(l){3-3}
$[0, 3)$ & 42.66 & 1.90 \\
$[3, 7)$ & 38.72 & 16.34 \\
$[7, + \infty)$ & 18.62 & 81.76 \\ \bottomrule
\end{tabularx}
\end{table}

\subsubsection{Variation in Sales}

\begin{figure}[H]
 \centering
 \includegraphics[width=\linewidth]{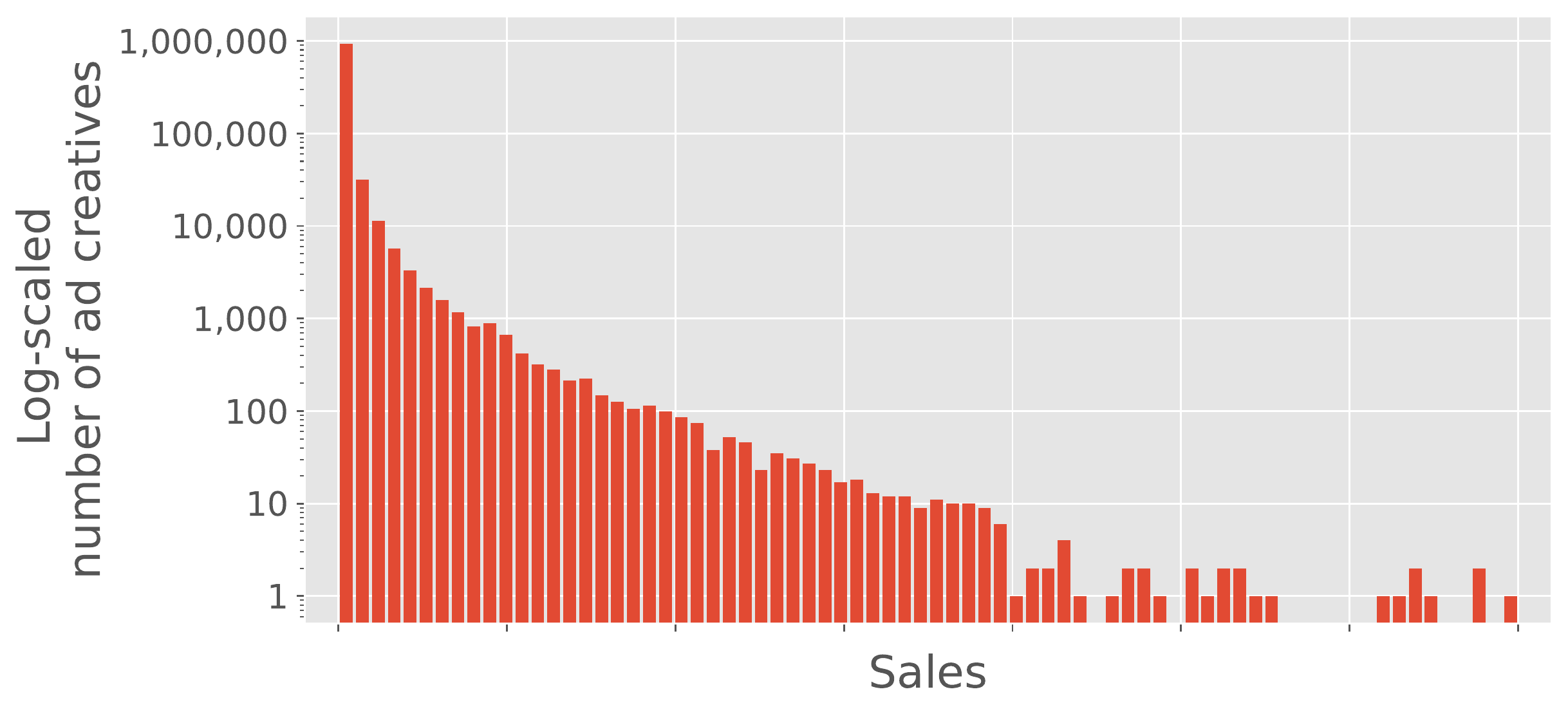}
 \caption{
 Histogram of ad creative sales in descending order.
 The y-axis is log-scaled.
 Most ad creatives are included in the first bin.
 Because the sales data are confidential, the actual values cannot be disclosed. 
 Thus, relative values are used here.
 }
 \label{fig:freq_of_sales}
\end{figure}

Figure \ref{fig:freq_of_sales} shows the histogram of ad creative sales in descending order.
Sales vary considerably for each ad creative.
Some are high-sales ad creatives, but most do not contribute to business revenue.
Thus, the impact of each ad creative on the revenue is very different.

Accordingly, we attempted to reflect the high-sales ad creatives for the prediction framework.
If the framework is trained with equal use of all ad creatives, the framework would reflect the trends of low-sales ad creatives excessively.
We assumed that the construction of the framework requires the weighting of learning depending on sales.
Therefore, we employ a CTR-weighting technique for the loss function, as described in Section \ref{sec:sales_based_loss_function}.

\subsection{Two Types of Discontinuation}\label{sec:analysis_two_types_of_the_discontinuation}
In this section, we discuss the differences between the two types of discontinuation based on the analysis in Section \ref{sec:analysis_of_the_dataset}.
We also describe how the differences can be handled effectively using our framework.

The short-term discontinuation is labeled \textit{cut-out}.
In digital advertising, a best practice is to create various patterns of ad creatives and look for effective ones by serving them.
Clearly, it is not easy to find an effective pattern, and thus, many ad creatives are discontinued in the short-term.
As Figure \ref{fig:freq_survival_day} shows, many ad creatives are discontinued early on, but this selection process is conducted manually.
Thus, it is a burden on the operation loads.
In fact, automation of the process has significant cost advantages (i.e., both time and money) from a business perspective.

In contrast, the discontinuation of an ad creative after it has been served for a long-term period is called \textit{wear-out} \cite{cornelia1988advertising}.
Wear-out occurs when the ad creatives are overexposed to customers and response drops.
The discontinuation of long-running ad creatives damages sales; however, the wear-out of ads is inevitable.
Thus, ad operators are tasked with creating new compelling ad creatives for their replacement.
Ad operators prioritize this task by estimating wear-out, which heavily depends on the operator's experience.
Thus, by predicting wear-out, we expect that the sales loss could be minimized.
In terms of ad operations, predicting wear-out has a different role from predicting cut-out.

In short, cut-out occurs when the user is not interested, and wear-out occurs when the user becomes fatigued.
The operations involved in each type of discontinuation also differ. 
In the case of a cut-out, the operator discontinues many inefficient ad creatives rapidly, whereas in the case of a wear-out, the operator estimates when an ad creative's performance degrades.
It is important that our framework predicts both types of discontinuation.
Predicting cut-out (for the short term) can save human effort and enable efficient advertising operations.
Predicting wear-out (for the long term) can have a significant impact on business revenue. 

The properties of these two types of discontinuation are so different that they should be considered as two different tasks. 
Therefore, we employ a two-term estimation technique (as described in Section \ref{sec:method_short_term_and_long_term_prediction}). Additionally, we analyze two real-world cases to evaluate the availability of each type of discontinuation (as described in Section \ref{sec:online}).

\section{Methodology}\label{sec:method}
We propose a multi-modal DNN-based framework for predicting the timing of the discontinuation of ad creatives, based on the idea of survival prediction.
First, we provide an overview of our framework.
Then, we introduce a loss function based on the hazard function for training our framework.
Additionally, we employ two techniques to enhance the performance for the prediction: (1) a two-term estimation technique with MTL and (2) a CTR-weighting technique for the loss function.
We believe that it is important to make the DNN model simpler, especially for business problems, and we introduce these effective learning techniques for the simple DNN model.

\subsection{Overview of the Proposed Framework}\label{sec:framework_overview}

\begin{figure}[H]
 \begin{adjustwidth}{-\extralength}{0cm}
 \centering
 \includegraphics[width=17cm]{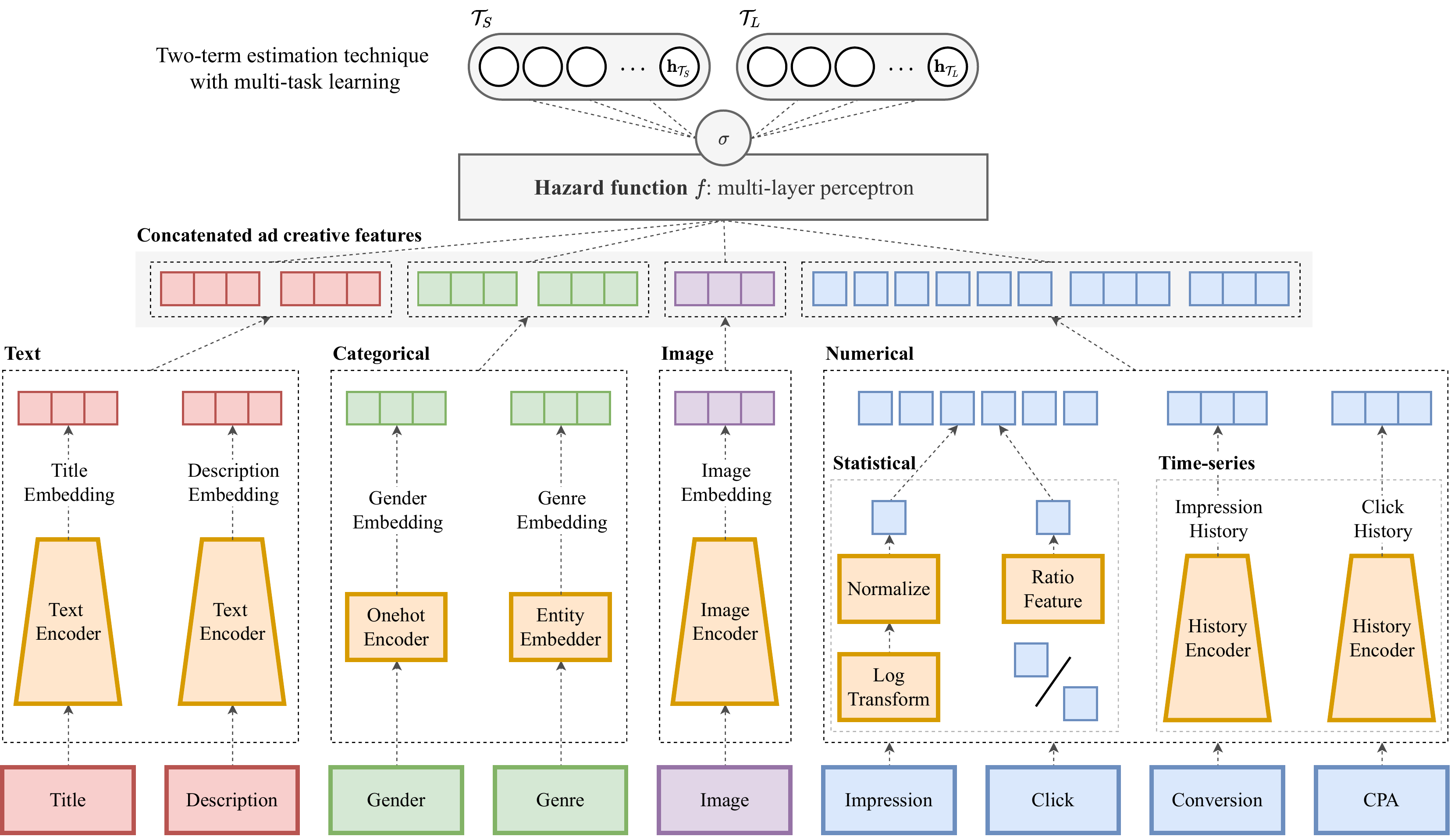}
 \end{adjustwidth}
 \caption{
 Outline of our framework that exploits a hazard function, which draws on the idea of survival prediction, to predict the discontinuation of ad creatives.
 The input includes the four types of features: text, categorical, image, and numerical features.
 The output is the hazard probability, which includes whether the target ad creative has been discontinued in each time interval.
 }
 \label{fig:proposed_framework}
\end{figure}

Figure \ref{fig:proposed_framework} shows the outline of our framework.
For predicting the timing of ad creative discontinuation, we adopt and integrate the hazard function, which is based on the discrete-time survival prediction strategy \cite{gensheimer2019scalable}.
From the input ad creative $\bm{x}$ observed in time $t$, our framework outputs hazard probability $\bm{h}$ for $L$ pre-defined time intervals $t$ such that $(0, t_1]$, $\mathcal{T} = (t_1, t_2]$, $\cdots$, $(t_{L-1}, t_L]$ through hazard function $f(\cdot|\cdot)$ with sigmoid function $\sigma$:
\begin{equation}
 \bm{h} = \sigma(f(t|\bm{x})) \in \mathbb{R}^L.
\end{equation}
The details of how the function is learned are in described in Section \ref{sec:learning_hazard_function}.

The input to our framework is ad creative $\bm{x}$ observed in time $t$; the framework uses the four types of features that constitute an ad creative: text features $\bm{x}_1$ (e.g., title, description), categorical features $\bm{x}_2$ (e.g., gender of the serving target, genre of the ad creative), image features $\bm{x}_3$ (e.g., the image attached to the ad creative), and numerical features $\bm{x}_4(t)$ (e.g., the number of impressions, clicks, conversions, and the CPA for the ad creative), which include statistical features and time-series features.
$\bm{x}_1$ to $\bm{x}_3$ are invariant with time interval $t$; $\bm{x}_4$ varies with $t$ due to the serving performance.

For the input features, each type of feature is encoded according to the type, such as text embeddings $\bm{e}_1$, categorical embeddings $\bm{e}_2$, image embeddings $\bm{e}_3$, and numerical embeddings $\bm{e}_4(t)$.
For all features, we performed conventional encoding and aggregation as well as encoding based on DNN techniques.
Due to the space limitation, the details of these input features and encoding processes are presented in Appendix \ref{sec:appendix_feature}.
Additionally, we briefly mention the properties of the features obtained by these encoding processes in the section.

For predicting the hazard probability, the encoded features are finally integrated using a multi-layer perceptron (MLP) to estimate the serving days of an ad creative in the pre-defined time intervals:
\begin{equation}
 \bm{h} = \sigma(f(t|\bm{x})) = \sigma(\mathrm{MLP}([\bm{e}_1, \bm{e}_2, \bm{e}_3, \bm{e}_{4}(t)])) \in \mathbb{R}^L.
\end{equation}

\subsection{Learning the Hazard Function for Ad Creative Discontinuation}\label{sec:learning_hazard_function}
We describe the problem settings of the hazard function in discrete-time survival prediction and how it can be learned for ad discontinuation.
The hazard function can model the main variable of interest as the time to an event.
To predict the timing of discontinuation, we are interested in discrete problems because currently discontinuation is done manually in real-world scenarios.

In discrete-time survival prediction, the follow-up time is broken up into time intervals, which are often right-closed and left-open.
Let $t'$ denote the time at which the event occurs, $l = 1, 2, \cdots, L$ index the time intervals, and $t_{l-1}$ and $t_{l}$ denote the lower and upper bounds of the $l$-th time interval, respectively.
The hazard probability within the $l$-th time interval is defined as follows:
\begin{equation}
 h_{l} = Pr(t' \in (t_{l-1}, t_l]|t' > t_{l-1}).
\end{equation}
Here, the time interval is the same setting defined in Section \ref{sec:framework_overview}.

The hazard probability can be modeled as a function of covariates to assess the effect of covariates on the time to the event of interest.
It is our belief that this property is appropriate for the discontinuation of ad creatives.
The hazard probability of the observation surviving the $l$-th time interval $h_{l}$ is equal to the $l$-th output of the hazard function, which is equivalent to the following: 
\begin{equation}
 h_{l} = \sigma(f_l(t_l|\bm{x})).
\end{equation}
where $f_l$ is the $l$-th output of hazard function $f(\cdot|\cdot)$.

The parameters are estimated through the maximization of the likelihood function for the discrete survival prediction strategy \cite{gensheimer2019scalable}.
The likelihood is written as follows:
\begin{equation}
 \mathcal{L}(\mathcal{T}, \bm{x}) = \prod_{l = 1}^{l'} h_{l}^{\delta_{l}}(1 - h_{l})^{(1- \delta_{l})}.
\end{equation}
where $l'$ is the number of time intervals that was observed, and $\delta_{l}$ is the event indicator within the $l$-th time interval, which is equal to 1 if the event occurs within that interval, and 0 otherwise.
To maximize the likelihood function, we minimized the following negative log-likelihood:
\begin{equation}
 \ell(\mathcal{T}, \bm{x}) = \sum_{l=1}^{l'} \delta_l \log h_l + (1 - \delta_l) \log (1-h_l).
\end{equation}
The full log-likelihood is the sum of the log-likelihood for each data and we minimize the log-likelihood by mini-batch gradient decent.

\subsection{Two-Term Estimation Technique: For Short- and Long-term Discontinuations}\label{sec:method_short_term_and_long_term_prediction}
We employ a two-term estimation technique to improve the performance for the two types of discontinuations (cut-out and wear-out).
To account for the difference between the two types of discontinuations, we train two prediction models by setting different time intervals for each model. 
We call these two models the \textit{short-term model} and the \textit{long-term model}, respectively.

The short-term model learns the cut-out features, and the long-term model learns the wear-out features.
In the short-term model, for time interval $\mathcal{T}_S$, which has $L_{\mathcal{T}_S}$ intervals, the target log-likelihood is defined as $\ell(\mathcal{T}_S, \bm{x})$.
In the long-term model, for time interval $\mathcal{T}_L$, which has $L_{\mathcal{T}_L}$ intervals, the target log-likelihood is defined as $\ell(\mathcal{T}_L, \bm{x})$.
Training with these different time intervals for each model is expected to enable learning of different properties of short- and long-term discontinuation, which will provide better predictions.

The MTL allows us to train short-term and long-term models in a unified single model.
In applying this technique to our framework to train the model, we used the following two likelihoods:
\begin{equation}
 \ell_{\mathrm{multi}}(\mathcal{T}_S, \mathcal{T}_L, \bm{x}) = \lambda \ell(\mathcal{T}_S, \bm{x}) + (1 - \lambda)\ell(\mathcal{T}_L, \bm{x}),
\end{equation}
where $\lambda$ is a regularization coefficient that controls the balance between the two loss functions.
In this study, we set $\lambda$ to 0.5.
As the fraction of short-term discontinuation is much larger than that of long-term discontinuation, tuning hyperparameter $\lambda$ would yield better performance.
The two hazard probabilities for the short- and long-term time intervals outputted from the MTL technique can be merged, multiplying by the last interval of the short-term time interval and the beginning of the long-term time intervals, to the last time interval of the long-term time intervals.
We used this result as the overall model of MTL with short- and long-term time intervals.

In particular, $\mathcal{T}_{S}$ and $\mathcal{T}_{L}$ were set to be as follows:
\begin{equation}
 \mathcal{T}_{S} = (1, 3], (3, 5], (5, 7], (7, 10], \quad \mathcal{T}_{L} = (1, 10], (10, 30], (30, 60], (60, 90], (90, 120].
\end{equation}
The time intervals for the short- and the long-term model were defined based on observations of a dataset in which most ads were discontinued in about 10 days or 120 days, respectively, under the assumption \cite{gensheimer2019scalable} that it was effective to make the interval width exponential.

\subsection{CTR-Weighting Technique for the Hazard-Based Loss Function}\label{sec:sales_based_loss_function}
 We employ a CTR-weighting technique for the hazard function-based loss function.
It is important to capture the features of ad creatives that have a high CTR, because this is known to contribute to business revenues, as stated in Section \ref{sec:analysis_of_the_dataset}.
Thus, we utilize the weighting technique to the loss function with the CTR of ad creatives:
\begin{equation}
 \ell_{\mathrm{CTR}}(\mathcal{T}, \bm{x}) = (r_{\mathrm{CTR}} + 1) \cdot \ell(\mathcal{T}, \bm{x}),
\end{equation}
where $r_{\mathrm{CTR}} = \#\texttt{click} / \#\texttt{impression}$ and generally takes the value $0 \le r_{\mathrm{CTR}} \le 1$.
Thus, we added one to this value to prevent the loss value from becoming too small for data with a low CTR.
For comparison, we used the impression-weighting technique $\ell_{\mathrm{imp}}$ with the same settings as the CTR-weighting technique. (Using sales directly as a weighting should be avoided in terms of business aspects because it would be too significantly affected by an advertiser's business.)

In the MTL, to apply our CTR-weighting technique, we used the following likelihood:
\begin{equation}
 \ell_{\mathrm{CTR}}^{\mathrm{multi}}(\mathcal{T}_S, \mathcal{T}_L, \bm{x}) = \lambda \ell_{\mathrm{CTR}}(\mathcal{T}_S, \bm{x}) + (1 - \lambda) \ell_{\mathrm{CTR}}(\mathcal{T}_L, \bm{x}).
\end{equation}

\section{Offline Experiments}\label{sec:offline}
In this section, we describe offline experiments with standard metrics using already-observed (i.e., offline) data. First, we explain the motivation and tasks of the offline experiments. Then, we describe the evaluation criteria of the experiment. The results of the evaluation according to the criteria will be presented and discussed.

\subsection{Motivation and Tasks of Offline Experiments}

The purpose of the offline experiments is to evaluate whether our framework can correctly predict ad discontinuation using observed data.
To this end, we used the real-world dataset of ad creatives shown in Section \ref{sec:dataset_overview}.
The discontinuation date of the ad creative was defined as the date on which the ad operator discontinued it manually.
The dataset was divided into a training set, a validation set, and a test set in a campaign-based and stratified fashion (In most advertising systems, advertisements are served in units of campaigns, and multiple creatives with similar trends are developed in a campaign. Thus, to avoid data leakage due to potential similarity, we divided the dataset depending on the campaigns), comprising 600,000, 200,000, and 200,000 ad creatives, respectively.

\subsection{Evaluation Criteria of Offline Experiments}

We evaluated our framework on the following aspects: (1) comparison with the conventional survival prediction-based methods and its suitable metrics, and (2) comparison with the classification and regression frameworks.
Regarding the former, considering our framework draws on the idea from survival prediction, we follow the evaluation in a survival prediction manner. 
For the latter, we evaluate the prediction made by the hazard function in our framework.

\subsubsection{Survival Prediction-Based Models with CI}

For the performance evaluation, we employed a CI \cite{harrell1982evaluating}, which is the standard measure for model assessment in survival prediction strategies.
A CI evaluates, for each random pair, the probability that the two predicted survival times will be in the same order as the actual survival time.
For context, CI = 0.5 is the average CI of a random prediction, whereas CI = 1.0 is the perfect ranking of a prediction.
We believe that this measure is sufficient to evaluate the framework based on our survival prediction.

From a business perspective, it is important to accurately predict the duration of top-selling creatives.
The differences between the sales of the ad creatives were found to be substantial, as shown in Section \ref{sec:dataset_overview}.
To determine whether the high-value creatives were accurately predicted, we evaluated only the top 25\% of the ad creatives in the dataset based on their sales values.

For comparison with the survival prediction technique, we used the conventional methods of the Cox time-varying proportional-hazards model (Cox time-varying PH) \cite{fisher1999time} and random survival forest \cite{ishwaran2008random}.
The hyperparameters of these methods were tuned by the validation set, and we report the prediction results for the test set.
To confirm the effects of our short-term and long-term models, we constructed an overall model that includes the time interval used in the short- and long-term models.
The hyperparameter settings of our framework are described in Appendix \ref{sec:hyper_parameter}.

\subsubsection{Classification and Regression Frameworks with F1 score}

To validate our framework, we compared it with binary classification and regression frameworks that simulate direct discontinuation. (Since both methods use the same model architecture (but with different loss functions), the computational costs are almost identical.)
This leads to an intuitive evaluation with other frameworks for which the CI cannot be directly applied.
The details of the classification and regression models are described in Appendix \ref{sec:classification_and_regression}.

We used the F1 score to compare our framework with the classification/regression frameworks.
The only difference between our framework and the others is the output objective; our framework predicts the hazard probability at each time interval in a single-task model, whereas the classification/regression framework predicts whether to discontinue at the time interval in each model.

\subsection{Evaluation Results of Offline Experiments}

In this section, we demonstrate the results of survival prediction-based models and classification/regression frameworks.

\subsubsection{Comparison of Survival Prediction-Based Models}

First, we compared the conventional method of the Cox time-varying PH and random survival forest in terms of the statistical and text features.
Table \ref{tab:result_concordance_index} shows in terms of the CI.
We can regard a model with the same features (i.e., the statistical and text features) under a single-task setting as a baseline based on our DNN.
Compared the DNN baseline with the conventional methods, our DNN-based framework showed better performance, with an approximate 10-point gain.
By using our framework, not only can the performance be improved, but various features, such as image and time-series features, can be accounted for as well.
Furthermore, comparing our DNN baseline with the full model of our framework (i.e., the model with all features under multi-task setting), we can confirm an average improvement of 3 points in prediction performance.

\begin{table}[H]
\centering
\caption{Comparison of concordance indexes with different time intervals and different features.
The time interval was compared for the short term, long term, and the whole period.
We calculated the concordance indexes using the prediction for all test data and the prediction for the top 25\% of sales data in the test data.}
\label{tab:result_concordance_index}
\begin{adjustwidth}{-\extralength}{0cm}
\begin{tabularx}{\fulllength}{llccccrrrrrr}
\toprule
\multicolumn{2}{c}{\multirow{5}{*}{\textbf{Model}}} & \multicolumn{4}{c}{\multirow{2}{*}{\textbf{Feature}}} & \multicolumn{6}{c}{\textbf{Concordance Index}} \\ \cmidrule(l){7-12} 
\multicolumn{2}{c}{} & \multicolumn{4}{c}{} & \multicolumn{2}{c}{\textbf{Short Term}} & \multicolumn{2}{c}{\textbf{Long Term}} & \multicolumn{2}{c}{\textbf{Overall}} \\ \cmidrule(r){3-6} \cmidrule(lr){7-8} \cmidrule(lr){9-10} \cmidrule(l){11-12} 
\multicolumn{2}{c}{} & \textbf{Stat.} & \textbf{Text} & \textbf{Image} & \textbf{Time.} & \multicolumn{1}{c}{\textbf{All}} & \multicolumn{1}{c}{\begin{tabular}[c]{@{}c@{}}\textbf{Top 25\%} \\ \textbf{of Sales}\end{tabular}} & \multicolumn{1}{c}{\textbf{All}} & \multicolumn{1}{c}{\begin{tabular}[c]{@{}c@{}}\textbf{Top 25\%} \\ \textbf{of Sales}\end{tabular}} & \multicolumn{1}{c}{\textbf{All}} & \multicolumn{1}{c}{\begin{tabular}[c]{@{}c@{}}\textbf{Top 25\%} \\ \textbf{of Sales}\end{tabular}} \\ \cmidrule(r){1-2} \cmidrule(lr){3-6} \cmidrule(lr){7-7} \cmidrule(lr){8-8} \cmidrule(lr){9-9} \cmidrule(lr){10-10} \cmidrule(lr){11-11} \cmidrule(l){12-12}
\multicolumn{2}{l}{Cox time-varying PH \cite{fisher1999time}} & \Checkmark & \Checkmark & & & 0.6098 & 0.7293 & 0.6574 & 0.6932 & 0.5287 & 0.5320 \\
\multicolumn{2}{l}{Random Survival Forest \cite{ishwaran2008random}} & \Checkmark & \Checkmark & & & 0.6213 & 0.7578 & 0.6832 & 0.7294 & 0.5311 & 0.5487 \\ \cmidrule(r){1-2} \cmidrule(lr){3-6} \cmidrule(lr){7-7} \cmidrule(lr){8-8} \cmidrule(lr){9-9} \cmidrule(lr){10-10} \cmidrule(lr){11-11} \cmidrule(l){12-12}
\textbf{Our framework} & Single-task & \Checkmark & \Checkmark & & & 0.7958 & 0.8262 & 0.7541 & 0.8001 & 0.5874 & 0.6001 \\
 & & \Checkmark & & \Checkmark & & 0.7962 & 0.8232 & 0.7536 & 0.8045 & 0.5536 & 0.6045 \\
 & & \Checkmark & & & \Checkmark & 0.7931 & 0.8346 & 0.7880 & 0.8397 & 0.5935 & 0.6101 \\
 & & \Checkmark & \Checkmark & \Checkmark & & 0.7962 & 0.8313 & 0.7575 & 0.8113 & 0.5908 & 0.6113 \\
 & & \Checkmark & \Checkmark & \Checkmark & \Checkmark & \textbf{0.8289} & \textbf{0.8640} & \textbf{0.7892} & \textbf{0.8456} & \textbf{0.6225} & \textbf{0.6456} \\ \cmidrule(r){2-2} \cmidrule(lr){3-6} \cmidrule(lr){7-7} \cmidrule(lr){8-8} \cmidrule(lr){9-9} \cmidrule(lr){10-10} \cmidrule(lr){11-11} \cmidrule(l){12-12} 
 & Multi-task & \Checkmark & \Checkmark & \Checkmark & \Checkmark & \textbf{0.8700} & \textbf{0.8712} & \textbf{0.9228} & \textbf{0.9293} & \textbf{0.7915} & \textbf{0.8049} \\ \bottomrule
\end{tabularx}
\end{adjustwidth}
\end{table}

For the short- and long-term models, the models that incorporated all features, such as image features, text features, and time-series features, had the best scores by an average of approximately 4 points.
In the short-term model, the text and image features contributed more compared with the long-term model.
However, in the long-term model, the time-series features contributed more than in the short-term model.
These differences are in line with the trends of the two types of discontinuation.
The effect of time-series data is described in Appendix \ref{sec:effect_of_timeseries_data}.

Compared with the models where the short- and long-term models were trained individually, we observed even better performance in the model that was trained simultaneously with MTL.
The performance improved significantly by approximately 5 points and 13 points in the short- and long-term models, respectively.
MTL further boosted the overall performance, with a clear improvement of about 17 points.

Table \ref{tab:ctr_weighted_loss} shows the MTL model performance comparison to confirm the effect of the CTR- and impression-weighting techniques.
By introducing the CTR-weighting technique, the prediction performance was improved in the short-term, long-term, and overall models.
The impression-weighting technique performed better than the vanilla model (e.g., without any weighting for the loss) but did not show a better performance compared with the CTR-weighting technique.

\begin{table}[H]
\caption{Comparison of prediction performance by proposed CTR-weighted loss in the multi-task model.
We compared the vanilla model (without any special loss), $\ell_{\mathrm{imp}}^{\mathrm{multi}}$, and $\ell_{\mathrm{CTR}}^{\mathrm{multi}}$.}
\label{tab:ctr_weighted_loss}
\newcolumntype{C}{>{\centering\arraybackslash}X}
\begin{tabularx}{\textwidth}{CR{3cm}R{3cm}R{3cm}}
\toprule
\multicolumn{1}{c}{\multirow{2}{*}{\vspace{-6pt}\textbf{Model}}} & \multicolumn{3}{c}{\textbf{Concordance Index}} \\ \cmidrule(l){2-4} 
\multicolumn{1}{c}{} & \multicolumn{1}{r}{\textbf{without}} & \multicolumn{1}{r}{\boldmath{$\ell_{\mathrm{imp}}^{\mathrm{multi}}$}} & \multicolumn{1}{r}{\boldmath{$\ell_{\mathrm{CTR}}^{\mathrm{multi}}$}} \\ \cmidrule(r){1-1} \cmidrule(lr){2-2} \cmidrule(lr){3-3} \cmidrule(l){4-4}
Short-term & 0.8700 & 0.8884 & \textbf{0.8958} \\ \cmidrule(r){1-1} \cmidrule(lr){2-2} \cmidrule(lr){3-3} \cmidrule(l){4-4}
Long-term & 0.9228 & 0.9297 & \textbf{0.9390} \\ \cmidrule(r){1-1} \cmidrule(lr){2-2} \cmidrule(lr){3-3} \cmidrule(l){4-4}
Overall & 0.7915 & 0.8060 & \textbf{0.8127} \\ \bottomrule
\end{tabularx}
\end{table}

In the short-term model with CTR weighting, the CI showed an improvement of approximately 3 points, and in the long-term model, the improvement was approximately 2 points.
The improvement in performance was observed even when impression weighting was introduced; however, the performance improved more when the CTR-weighting technique was introduced.
We assume that the large variance in the number of impressions is the reason for the difference between these scores.
Thus, weighting with CTR enables training for more accurate prediction.

\subsubsection{Comparison of Classification and Regression Frameworks}
Table \ref{tab:bc_vs_reg_vs_sa} compares our framework and classification/regression frameworks.
Compared with other frameworks, our framework performed better by approximately 40 points in the short-term model and by 60 points in the long-term model.
In the short-term model, our framework achieved superior performance, about twice that of the other frameworks.
In the long-term model, the classification/regression frameworks did not appear to perform well.
Based on these results, our hazard function-based approach represents an effective method for predicting ad creative discontinuation problems.

\begin{table}[H]
\caption{Comparison of the F1 score in discontinuation prediction after $N$ days using the binary classification framework and regression framework.}
\label{tab:bc_vs_reg_vs_sa}
\newcolumntype{C}{>{\centering\arraybackslash}X}
\begin{tabularx}{\textwidth}{lR{2cm}R{2cm}R{2cm}R{2cm}}
\toprule
\multicolumn{1}{c}{} & \multicolumn{2}{c}{\textbf{Short-Term}} & \multicolumn{2}{c}{\textbf{Long-Term}} \\ \cmidrule(r){2-3} \cmidrule(l){4-5}
\multicolumn{1}{c}{} & \multicolumn{1}{r}{\textbf{3 Days}} & \multicolumn{1}{r}{\textbf{7 Days}} & \multicolumn{1}{r}{\textbf{30 Days}} & \multicolumn{1}{r}{\textbf{90 Days}} \\ \cmidrule(r){1-1} \cmidrule(lr){2-2} \cmidrule(lr){3-3} \cmidrule(lr){4-4} \cmidrule(l){5-5}
Classification framework & 0.3208 & 0.3106 & 0.1098 & 0.0980 \\
Regression framework & 0.3813 & 0.2852 & 0.0969 & 0.0735 \\ \cmidrule(r){1-1} \cmidrule(lr){2-2} \cmidrule(lr){3-3} \cmidrule(lr){4-4} \cmidrule(l){5-5}
\textbf{Our framework} & \textbf{0.7988} & \textbf{0.7796} & \textbf{0.7287} & \textbf{0.7004} \\ \bottomrule
\end{tabularx}
\end{table}

\subsection{Discussion of Offline Experiments}

We discuss the following three points: (1) the effect of input features on prediction performance, (2) the effect of two-term estimation with MTL and CTR-weighting Technique, and (3) the comparison of our hazard function-based framework and classification/regression framework.

\subsubsection{The Effect of Input Features on Prediction Performance}

In our framework, the effective feature of the prediction of short- and long-term ad discontinuation is different. In the short-term model under the single-task setting, the text and image features that make up the ad creative contribute to the prediction. The short-term discontinuation (i.e., cut-out) may be caused by many users who are not immediately interested in the ad creatives. Since the text and image affect the user's impression of the ad creative, this result reflects characteristics of the short-term discontinuation. In the long-term model, the time-series feature contribute to the prediction. The long-term discontinuation (i.e., wear-out) is caused by the user getting fatigued, as described in Section {\ref{sec:dataset_overview}}. This is due to the fact that users have been repeatedly interacting with similar or the same ads, and thus the CPA has been decreasing. Since the time-series feature captures the user's fatigue by change of the ad performance, this result reflects characteristics of the long-term discontinuation. As a result, we confirmed that each input feature contributes to the prediction of both cut-out and wear-out appropriately.

\subsubsection{The Effect of Two-Term Estimation with MTL and CTR-Weighting Technique}

Ad creative discontinuations are of two types with different characteristics, as described in Section {\ref{sec:dataset_overview}}. To accurately predict the appropriate timing for the discontinuations, we introduce MTL into our two-term estimation technique. Additionally, we introduce the CTR-weighting technique to learn more valuable ad creative features. We confirmed that these techniques contribute to the prediction of ad creative discontinuation.

For ad creative discontinuation that has two different types of characteristics, we demonstrated that the model with our two-term estimation technique is superior to the single (overall) model. The short- and long-term models trained by the two-term estimation technique achieved much higher prediction performance than the single overall model, which included both short- and long-term time intervals. Furthermore, we found that training a unified model of these two-term models by MTL achieved better performance than the short- and long-term models alone. We believe that the unified model with MTL can learn the fine-grained property of short- and long-term discontinuation than the single overall model.

Our CTR-weighting technique is expected to learn more valuable ad creative features accurately. The results showed that our model was a more accurate predictor with CTR-weighting loss than without it. It improved the accuracy of the prediction of discontinuation of top-selling ad creatives and the overall accuracy. Consequently, it can be assumed that accurate training of effective ad creatives will have a positive effect on the overall prediction. 

\subsubsection{Our Hazard Function-Based Framework vs. Classification/Regression Framework}

Our framework achieves much higher prediction performance for the ad discontinuation prediction problem by introducing the hazard function instead of the conventional classification/regression framework. We assume that there are two reasons due to the property of the discontinuation problem: (1) the data imbalance and (2) the time dependence of the discontinuation. 
Regarding the data imbalance, most of the ad creatives were discontinued early. If the classification/regression framework is trained on such data, it may output only biased predictions. 
Regarding the time dependence of the discontinuation, the probability of the ad creative discontinuation increases as time goes on. The classification/regression framework does not have the assumption of time dependency, while the hazard function-based framework is known to have the assumption. 

We consider that the hazard function appropriately captures the property of the two reasons derived from the ad creative discontinuation problem. 
In contrast, the classification/regression framework does not seem to sufficiently handle the problems of the data imbalance and the time dependence. 
In particular, the classification- and regression-based long-term models may output only discontinuation, which may be why the F1 score is low. 
Furthermore, as ad creatives tend to be discontinued due to their decreasing effectiveness of serving over time, our framework, which can consider the discontinuation probability at the previous time interval, is considered to function effectively.

\section{Case Studies}\label{sec:online}
\subsection{Motivation and Tasks of the Case Studies}\label{sec:online_motivation_and_tasks}

As our goal is to support ad operations (specifically ad discontinuation), we needed to verify how effective our framework is in practical cases.
In Section \ref{sec:offline}, we confirmed that our framework performs well for predicting the ad creative discontinuation.
However, we should evaluate with other criteria in addition to the traditional performance criteria in survival prediction such as CI.

In this section, we evaluate the availability of our framework on two real-world cases according to the following research questions using as yet unobserved data:
\begin{itemize}
 \item RQ1: How does the efficiency of our framework for discontinued ad creatives compare to that of human operation?
 \item RQ2: How close is the order of the ad discontinuation based on the framework's predictions to the actual manual order?
\end{itemize}

In the first case, we evaluate the cut-out (i.e., short-term) performance of our framework.
Our purpose in the short-term is to reduce the operation loads on the operators from having to discontinue many ad creatives manually.
We confirm how the efficiency changes when the ad is discontinued according to our framework, compared with that under manual operation.
In the second case, we evaluate the wear-out (i.e., long-term) performance.
Ad operators must prioritize tasks that create new compelling ad creatives to replace of long-running ads, as described in Section \ref{sec:analysis_two_types_of_the_discontinuation}.
Therefore, it is crucial to identify an appropriate order for ad discontinuation in the ad business.

For the evaluation, we use the best model identified in Section \ref{sec:offline}.
When the hazard probability from the model exceeds 0.9 for the first time, we consider that the ad creative has been discontinued in the time interval.
The two practical cases are analyzed for another period to confirm the time robustness with 400,000 ad creatives for three months, unlike in the offline experiments. (These data were sampled randomly to ensure confidentiality.)

\subsection{Evaluation Criteria of the Case Studies}

To evaluate our framework more closely against real-world ad operations, we conducted case studies in terms of CPA (While we understand the importance of the volume metric of the ads, we keep it private for business reasons because the metric is closely linked to real-world commercial systems.) and the order in which the ad creative is discontinued. The evaluation results with the criteria respond to the aforementioned research questions. 
We describe the details of these criteria below.

\subsubsection{CPA Ratio: The Ratio of the Actual CPA to the Target CPA}\label{sec:evaluation_metrics_cpa_ratio}

We introduce practical metrics to assess the performance of ad creatives.
One of the major metrics for deciding discontinuation in ad operation is the \textit{CPA ratio}, which measures the efficiency of the CPA.
We use this metric as the baseline for the experiments.
The CPA ratio is defined as follows:%
\begin{equation}
 \texttt{CPA ratio} = \texttt{Actual CPA} / \texttt{Target CPA}.
\end{equation}
The actual CPA is calculated based on the number of sales and conversions and is the indicator of how much the ad creative costs compared with the conversions.
The target CPA is the CPA that the advertiser wants.
This metric measures CPA efficiency, which represents both the costs and volume. We should avoid simply evaluating a low CPA because it would reduce the exposure of the ad when the CPA is low. Advertisers are expected to minimize the volume within their budgets. We believe that the CPA ratio can be used to evaluate both the cost and volume.

The CPA ratio should not be high but not low either.
When the metric is low, the setting of the cost can be satisfied, but the volume can be increased by allowing a higher actual CPA.
The closer the CPA ratio is to 1, the more profitable it is for the advertiser and the media.
This metric is actually used by ad operators to decide whether discontinue ad creatives.
We assume that evaluation based on this metric is similar to professional operator decisions.

\subsubsection{The Order of Ad Discontinuation}
We evaluated the order of discontinuation for RQ2.
To prepare for the sudden discontinuation of long-running ad creatives, it is necessary to ascertain to evaluate whether our framework can accurately predict the order in which ads are discontinued.
To evaluate the order of ad discontinuation, we compared the normalized discounted cumulative gain (NDCG) \cite{jarvelin2002cumulated}.
The NDCG is calculated for ad creatives in the order in which they are discontinued by the ad operator and in the order in which our framework predicts discontinuations.

\subsection{Evaluation Results of the Case Studies}

In this section, we demonstrate the results of CPA ratios and discontinuation-order evaluation. The results suggest that our framework could support human operators.

\subsubsection{Comparison of CPA Ratios}
To answer RQ1, we evaluated the CPA efficiency if the ad operators decide to discontinue ad creatives according to our framework.
Our framework predicted the discontinuation date based on only one day of data from serving.
The CPA ratios of our framework are calculated based on the predicted discontinuation date, and the ratios for human performance are calculated based on the actual discontinuation date.
We note that the discontinuations by human decision used the performance after the fist day, contrary to our framework.
This experiment assumed the cut-out case, and thus, we evaluated the output of the short-term model, as described in Section \ref{sec:online_motivation_and_tasks}.

Table \ref{tab:practical_short_term} shows the comparison of the CPA ratios.
Compared to the best model under the single-task setting, the best model under the multi-task setting successfully predicted the discontinuation with a CPA ratio close to one. Furthermore, the multi-task model shows better performance of discontinuation in terms of the CPA ratio than the human operator's decision.
Overall, we confirmed that the single- and multi-task models have the same level of human performance.

\begin{table}[H]
\caption{Comparison of the CPA ratios for predicting cut-out in the practical evaluation experiment.
The CPA ratio is calculated as $\texttt{Actual CPA}/ \texttt{Acceptable CPA}$.
The closer the CPA ratio is to 1, the more profitable it is for the advertiser and the media.
Human performance is calculated on the day the ad operator discontinues the ad creative.
}
\label{tab:practical_short_term}
\newcolumntype{C}{>{\centering\arraybackslash}X}
\begin{tabularx}{\textwidth}{CCCC}
\toprule
\textbf{Short-Term} & \textbf{Single-Task} & \textbf{Multi-Task} & \textbf{Human Performance} \\ \cmidrule(l){1-1} \cmidrule(lr){2-2} \cmidrule(lr){3-3} \cmidrule(r){4-4}
CPA ratio & {1.13 $\pm$ 0.92} &{\textbf{1.12 $\pm$ 0.78}} & {1.18 $\pm$ 0.46} \\ \bottomrule
\end{tabularx}
\end{table}

\subsubsection{Comparison of the Discontinuation-Order Evaluation}
To answer RQ2, we determined how close the discontinuation order predicted by our framework was to the actual discontinuation order.
Correctly predicting which ads are likely to discontinue allows ad operators to set operational priorities.
This experiment assumed the wear-out case; hence we evaluated the output of the long-term model, as described in Section \ref{sec:online_motivation_and_tasks}.

To decide on a discontinuation order using our framework, we utilized the predicted discontinuation date and the hazard probability.
To predict the wear-out case, our framework used only the first 10 days of data from serving.
In each time interval, we sorted ad creatives based on the probability, and we concatenated the ordered ad creatives in each time interval.

To compare the evaluation results, we introduced two practical rule-based indicators, which are similar to practical operations.
One was to sort by sales, namely, \textit{sales order}.
The business is seriously affected if an ad with high sales is discontinued.
The other was to sort by the CPA ratio, namely, the \textit{CPA ratio order}, because ads with a high CPA ratio are easy to discontinue.
These indicators are considered to be correlated with human decision-making. (Ad operators usually monitor these indicators daily.)

\begin{figure}[H]
 \centering
 \includegraphics[width=0.6\linewidth]{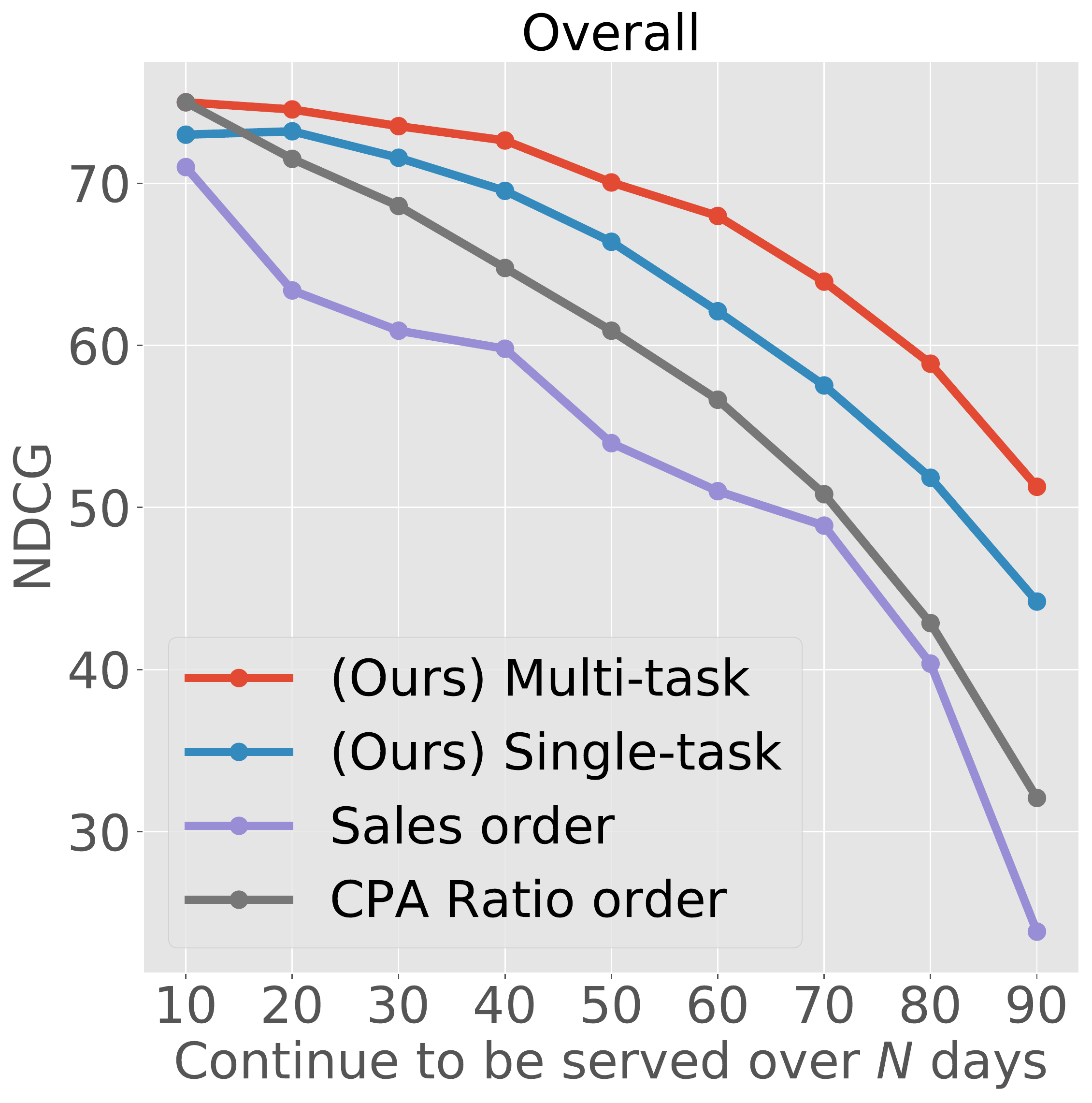}
 \caption{
 Comparison of the NDCG for the following prediction and actual discontinuation orders for each continuous serving days: by our framework (i.e., single-task and multi-task) and by the rule-based on the sales and the CPA ratio.
 }
 \label{fig:long_term_cpa_ratio}
\end{figure}

Figure \ref{fig:long_term_cpa_ratio} shows the NDCG of the prediction of our framework and the actual discontinuation orders for each continuous serving day.
This experiment targeted ad creatives that have been served for more than 10 days.
Each evaluation was conducted every 10 days for ad creatives within continuous serving days.
For example, when the x-axis value was 20, the y-axis showed the NDCG value with ad creatives with 20 or more days of continuous serving.
This evaluation determined how accurately we could predict future discontinuation.

\subsection{Discussion of the Case Studies}

We discuss the results of the case study in terms of the CPA ratio and the order of ad discontinuation. These lead to the answers to RQ1 and RQ2, respectively.

\subsubsection{RQ1: Perspective of CPA ratio}

In Table {\ref{tab:practical_short_term}}, we confirmed that both single- and multi-task models outperform the performance of the human operator. Because the performance of our models does not use unseen future data, which are used in the human performance, we posit that the same level of the performance is sufficient. This result suggests that, even if we decide the discontinuation using the prediction of our framework, the CPA efficiency does not get worse.

Clearly, the performance after ad discontinuation cannot be known, and thus, when the prediction result comes after the actual discontinuation date, the CPA ratio is the same as the actual CPA ratio.
We posit that this problem does not affect our purpose, as our proposal is not to outperform but rather support human decision-making.

\begin{quotation}
\textit{To answer {\bf RQ1}, when our framework discontinues an ad creative, the CPA efficiency is almost equal to human operation.}
\end{quotation}

\subsubsection{RQ2: Perspective of the Order of Ad Discontinuation}

Overall, our framework provided appropriate predictions of the discontinuation order compared with sales- and CPA ratio-based metrics. The CPA ratio and our framework exhibited the same performance for the first 10 days. However, when the number of serving days was increased, our framework performed better. In other words, the CPA ratio is suitable indicator for predicting which ad will be discontinued in the near future but cannot predict the discontinuation days in the long term, unlike our framework. Because it is important to plan priorities using long-term predictions, our framework plays an important role in prioritizing ad operations.

\begin{quotation}
\textit{To answer {\bf RQ2}, our framework achieves a closer prediction of the order of the actual discontinuation than practical indicators. }
\end{quotation}

\section{Limitation and Future Work}\label{sec:limitation}
Our framework has been deployed in our production environment, and we engage in regular discussions with the operations team to figure out ways to further improve the operation efficiency.
Throughout the operation with the proposed framework, we found some issues/limitations related to the model itself and operation using our framework. One issue with the model is that it requires some experience in setting the appropriate time interval for the two-term estimation. In the experiments, we determined time intervals that were empirically known to perform well. The remaining issue in the operation is how to set the threshold for the discontinuation against the predicted probability by the framework. For these issues, we must work with experienced operators to work out the final operational approach.
In the future, by analyzing the prediction results, we intend to establish the best practice to utilize the result and publish our framework for customers.
Additionally, although this may be one way to address the above limitation, we intend to train a model that rewards the operator's discontinuation behavior based on reinforcement learning.

\section{Conclusions}\label{sec:conclusion}
Ad creative discontinuation has a major role in ad operations.
Although ad operation efficiency using ML methods is an active research topic, there have been few practical studies regarding how to predict the appropriate timing of ad creative discontinuation.
To achieve this goal in a real-world environment, we first analyze 1,000,000 real-world ad creatives and found that there are two types/patterns of discontinuation. Based on the finding, we proposed a multi-modal DNN-based framework for predicting the timing of the discontinuation of ad creatives that draws on the idea of survival prediction.
Our framework has a loss function based on the hazard function and contains the following two simple but novel techniques: (1) a two-term estimation technique with multi-task learning and (2) a CTR-weighting technique for the loss function.

Evaluations using the large-scale ad creatives confirmed that our two-term estimation technique significantly improves the prediction performance, and CTR-weighting technique further improved the performance.
Compared with classification/regression frameworks, we observed that our framework performed notably better for predicting discontinuation.
Additionally, for the two practical cases in the case studies, our framework achieved equivalent CPA efficiency with manual operation for short-term prediction and a higher NDCG with manual operation compared with other indicators using actual sales metrics (sales order, CPA ratio order) in long-term predictions.
While the DNN-based architecture in our framework is not entirely novel, we addressed a crucial problem encountered in real-world ad operations and confirmed its feasibility in practical cases.

\vspace{6pt}

\authorcontributions{Conceptualization, H.I. and Y.S.; Data curation, Y.S. and S.K.; Formal analysis, S.K. and Y.S.; Investigation, S.K. and Y.S.; Methodology, S.K. and Y.S.; Project administration, Y.S.; Software, S.K.; Supervision, H.I. and Y.S.; Validation, S.K., H.I. and Y.S.; Visualization, S.K.; Writing---original draft, S.K. and Y.S.; Writing---review and editing, H.I. and Y.S. All authors have read and agreed to the published version of the manuscript.}

\funding{This work was partially supported by JSPS KAKENHI under Grant 21J14143.}

\institutionalreview{Not applicable.}

\informedconsent{Not applicable.}

\dataavailability{Not applicable.}

\acknowledgments{We thank the Gunosy Ad Engineering team for their useful feedback and Hironori Yamamoto for helping us with the deployment of the proposed framework.}

\conflictsofinterest{The authors declare no conflict of interest.}

\appendixtitles{yes} %
\appendixstart
\appendix

\section{Detail of Features}\label{sec:appendix_feature}
Our framework used the four feature areas that constitute the ad creative: text features, categorical features, image features, and numerical features.
The details of these input features and encoding processes are described as follows.

\subsection{Text Features}
Text features were extracted from the title and description of an ad creative.
Specifically, words were embedded from the title and the description and encoded by a text encoder using a recurrent neural network (RNN)-based model.
For the RNN-based model, we used bi-directional long short-term memory (LSTM) \cite{hochreiter1997long} in our implementation.
We can expect the extracted text features to consider the context of the ad creative text.
Subsequently, title embedding and description embedding were calculated as the last hidden state of each text encoder as: $\bm{h}^{\mathrm{title}} \in \mathbb{R}^{d_{\mathrm{title}}}$ and $\bm{h}^{\mathrm{desc}} \in \mathbb{R}^{d_{\mathrm{desc}}}$.
Finally, we concatenated these embeddings to text embeddings:
\begin{equation}
 \bm{e}_1 = [\bm{h}^{\mathrm{title}}; \bm{h}^{\mathrm{desc}}] \in \mathbb{R}^{d_{\mathrm{title}} + d_{\mathrm{text}}}.
\end{equation}

\subsection{Categorical Features}
Categorical features were extracted from the gender of the serving target and the genre of the ad creative.
Specifically, the gender of the distribution target was extracted as $\bm{x}^{\mathrm{gender}} \in \mathbb{R}^{d_{\mathrm{gender}}}$, which was one-hot encoded.
Additionally, the genre of the ad creative, which was embedded by entity embedding method \cite{guo2016entity}, was extracted as $\bm{x}^{\mathrm{genre}} \in \mathbb{R}^{d_{\mathrm{genre}}}$.
We can expect the extracted categorical features to consider the demographic information of the target users.
Finally, we combined these embeddings in categorical embeddings:
\begin{equation}
 \bm{e}_2 = [\bm{x}^{\mathrm{gender}}; \bm{x}^{\mathrm{genre}}] \in \mathbb{R}^{d_{\mathrm{gender}} + d_{\mathrm{genre}}}.
\end{equation}

\subsection{Image Features}
Image features were extracted from an image of an ad creative.
Specifically, image embedding $\bm{e}_3$ was obtained by encoding $d_{h} \times d_{w}$ sized ad image $\bm{x}_3$ using the ResNet34 \cite{he2016deep} pretrained by ImageNet:
\begin{equation}
 \bm{e}_3 = \mathrm{ResNet34}(\bm{x}_3).
\end{equation}
We can expect the extracted image feature to consider the characteristics of the ad image.

\subsection{Numerical Features}
Numerical features include the number of impressions, clicks, and conversions and the CPA for an ad creative. 
The numerical features consist of statistical and time-series features.

The statistical feature is composed of statistical information about target date $t$.
The CTR was obtained from the number of impressions and the number of clicks, and the CVR was obtained from the number of clicks and the number of conversions.
We used the following features as the statistical feature: the number of impressions $x_{s1}$, clicks $x_{s2}(t)$, conversions $x_{s3}(t)$, the CTR $x_{s4}(t)$, the CVR $x_{s5}(t)$, and the CPA $x_{s6}(t)$.
These features were normalized through logarithmic transformation.
Finally, we concatenated these features as statistical embeddings:
\begin{equation}
 \bm{e}_4^{\mathrm{Stat}}(t) = [x_{s1}(t);  x_{s2}(t);  x_{s3}(t);  x_{s4}(t);  x_{s5}(t);  x_{s6}(t)].
\end{equation}

Time-series features were extracted from impressions and clicks accumulated from the first day of serving to the target date $t$ based on statistical features.
Specifically, from the accumulated impressions and clicks, those encoded by the history encoder with different LSTM models were obtained as an impression history embedding $\bm{h}^{\mathrm{imp}}$ and a click history embedding $\bm{h}^{\mathrm{click}}$, respectively.
These embeddings are the last hidden states encoded by the history encoder of the serving performance data up to time $t$.
Then, we combined these features into time-series embeddings:
\begin{equation}
 \bm{e}_4^{\mathrm{Time}}(t) = [\bm{h}^{\mathrm{imp}}; \bm{h}^{\mathrm{click}}] \in \mathbb{R}^{d_{\mathrm{imp}} + d_{\mathrm{click}}}.
\end{equation}
Finally, we combined statistical embedding $\bm{e}_4^{\mathrm{Stat}}(t)$ and time-series embedding $\bm{e}_4^{\mathrm{Time}}(t)$ into numerical embeddings:
\begin{equation}
 \bm{e}_4(t) = [\bm{e}_4^{\mathrm{Stat}}(t);  \bm{e}_4^{\mathrm{Time}}(t)].
\end{equation}
We can expect the extracted numerical features to consider the both statistical and time-series properties of the ad creative.

\section{Implementation Details}\label{sec:hyper_parameter}
The texts of the ad creatives, written in Japanese, were split into words using MeCab \cite{kudo2006mecab}, which is a type of morphological analysis for Japanese texts.
The custom dictionary mecab-ipadic-neologd (\url{https://github.com/neologd/mecab-ipadic-neologd}), which includes various neologisms, was used as the dictionary.

We used pre-trained word2vec \cite{suzuki2018joint} to initialize the word embedding layer in our framework.
For the text encoder, $d_{\mathrm{title}}$ and $d_{\mathrm{desc}}$ were set to 16.
For the history encoder, $d_{\mathrm{imp}}$ and $d_{\mathrm{click}}$ were set to 10.
For an image encoder, we extracted 512 dimensional feature vectors.
The feature vectors were then converted using a shallow MLP to $d_{\mathrm{img}} = 16$ dimensional image features.

We used pycox (\url{https://github.com/havakv/pycox}) to implement the base of our survival prediction framework.
The mini-batch size was set to be 32, and the number of epochs was set to be 50.
Adam \cite{kingma2014adam} was used for the model optimization.

In all experiments, we performed the evaluation on the test set only once.
The experiments were conducted on an Ubuntu PC with one GeForce GTX 1080 Ti GPU.
As our framework is relatively small, we ran the case studies on the CPU.

Regarding the deployment of our framework, we dockerized the implementation of our framework to work on Amazon Web Service (AWS).
The prediction from our framework is output daily to a Google spreadsheet for discussion with the ad operation team.

\section{Classification and Regression Frameworks}\label{sec:classification_and_regression}
As binary classification models, we built models that predicted discontinuation after 3, 7, 30, and 90 days.
We evaluated the predictions of the model specific to each time interval.
In training the binary classification model, we used the binary cross entropy loss to train the model.
For a fair comparison, we applied a CTR-weighting technique based on our idea to the comparison models.

Regarding the regression model, we built models for short-term and long-term.
The regression models were trained on data with time intervals adjusted for short- and long-term intervals.
For training the regression model, we used mean squared error loss to train the model.
We used the time interval that was closest to the prediction made by the regression model as the final prediction.
Similar to binary classification, we applied the CTR-weighting technique for a fair comparison.

The same architecture as our single-task model (as described in Section \ref{sec:offline}) was used in these models, and the same features were used as inputs.
Note here that we considered an evaluation based on the mean squared error, but we believe that the metrics and results presented in Section \ref{sec:offline} are more suitable for comparison.

\section{Effects of Time-Series Data}\label{sec:effect_of_timeseries_data}
\begin{figure}[H]
 \begin{adjustwidth}{-\extralength}{0cm}
 \centering
 \begin{minipage}{0.26\linewidth}
 \includegraphics[height=6cm]{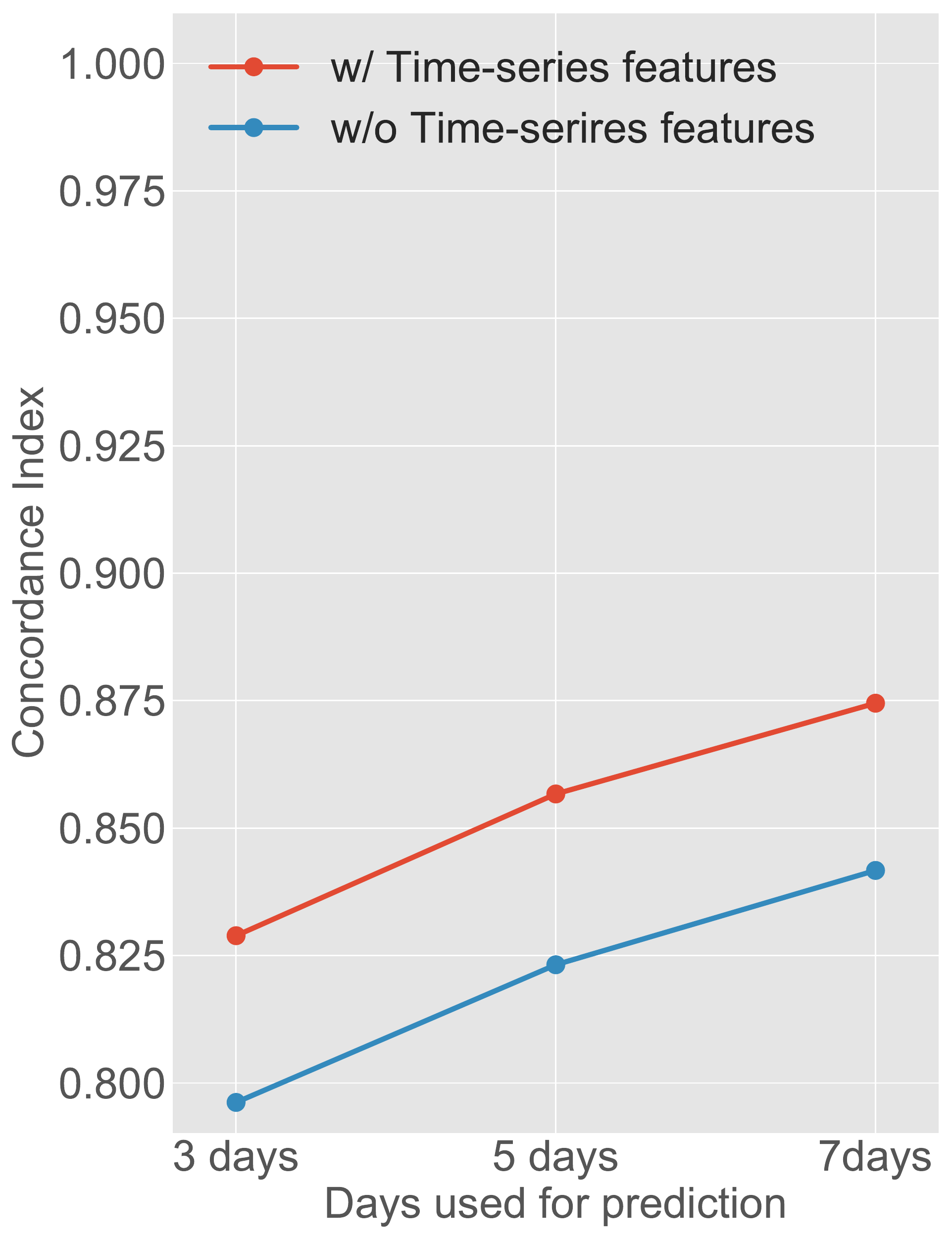}
 \subcaption{Short-term}
 \label{fig:timeseries_short}
 \end{minipage}
 \begin{minipage}{0.7\linewidth}
 \includegraphics[height=6cm]{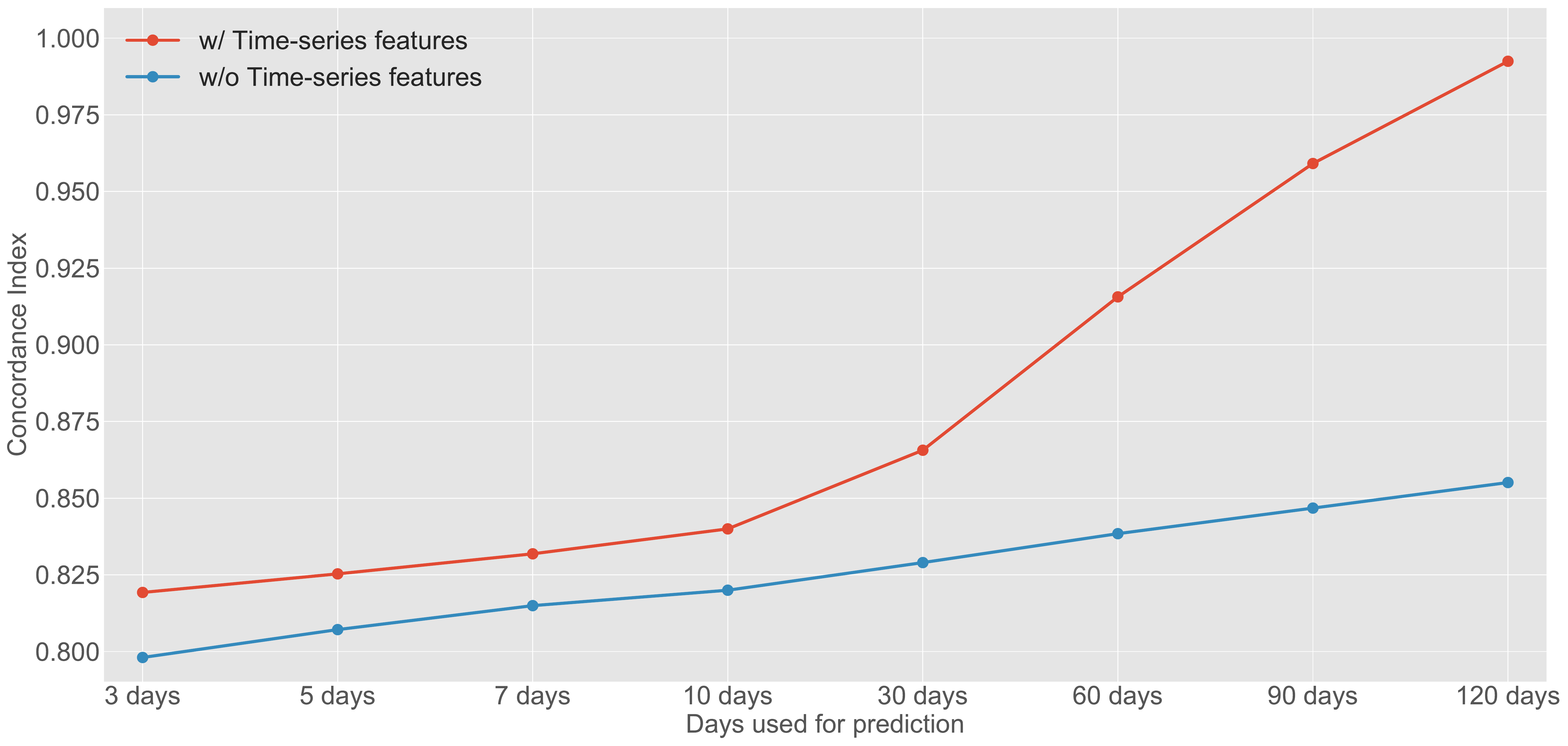}
 \subcaption{Long-term}
 \label{fig:timeseries_long}
 \end{minipage}
 \end{adjustwidth}
 \caption{
 Comparison of the effects of time-series features with daily changes in ad discontinuation prediction task.
 For the short- and the long-term models, the more days that can be used for prediction, the higher the performance.
 In the long-term model, when used for prediction for 30 days or more, the concordance index is dramatically improved compared with the case where the time-series feature is not used.
 }
 \label{fig:timeseries}
\end{figure}

Figure \ref{fig:timeseries} shows the effect on prediction performance of the time-series features, changing the number of days used for prediction.
Figure \ref{fig:timeseries_short} shows the change in the CI in the short-term model.
As described above, the model to which the time-series feature is input achieves a higher score compared with the model to which no time-series feature is input.
Additionally, as the number of days used for prediction increased, the prediction performance of each model improved.
Figure \ref{fig:timeseries_long} shows the change in the CI in the long-term model.
Similar to the short-term model, in the long-term setting, the model with time-series features has a higher score compared with the model without time-series features.
Therefore, it appears that the effectiveness of the time-series features for prediction increases with time.

\begin{adjustwidth}{-\extralength}{0cm}

\reftitle{References}

\end{adjustwidth}
\end{document}